\documentclass{webofc}
\usepackage[varg]{txfonts}   
\usepackage{xspace} 
\def\fb   {\ensuremath{\aunit{fb}}\xspace}
\def\invfb   {\ensuremath{\fb^{-1}}\xspace}

\def\lhc    {\mbox{LHC}\xspace}
\newcommand{\aunit}[1]{\ensuremath{\text{\,#1}}} 
\newcommand{\gevcc}{\ensuremath{\aunit{Ge\kern -0.1em V\!/}c^2}\xspace}
\newcommand{\gevc}{\ensuremath{\aunit{Ge\kern -0.1em V\!/}c}\xspace}
\def\sqsnn {\ensuremath{\protect\sqrt{s_{\scriptscriptstyle\text{NN}}}}\xspace}
\def\nb {\aunit{nb}\xspace}
\def\invnb {\ensuremath{\nb^{-1}}\xspace}
\newcommand{\tev}{\aunit{Te\kern -0.1em V}\xspace}

\def\Zj{\ensuremath{\Z j}\xspace}
\def\Zc{\ensuremath{\Z c}\xspace}
\def\Zcj{\ensuremath{\mathcal{R}^c\kern-0.4em{\raisebox{-0.2em}{$\scriptstyle j$}}}\xspace}

\def\to                 {\ensuremath{\rightarrow}\xspace}
\def\Pmu         {\ensuremath{\mu}\xspace}
\def\mumu       {{\ensuremath{\Pmu^+\Pmu^-}}\xspace}

\def\PZ  {\ensuremath{\mathrm{Z}}\xspace} 
\def\Z      {{\ensuremath{\PZ}}\xspace}

\def\gevcsquare    {\ensuremath{{\,(\mathrm{Ge\kern -0.1em V\!/}c)^2}}\xspace}
\def\mygevc        {\ensuremath{{\mathrm{Ge\kern -0.1em V\!/}c}}\xspace}
\def\unitgev   {\ensuremath{\mathrm{[Ge\kern -0.1em V]}}\xspace}

\def\pPb   {\ensuremath{p\xspace\mathrm{Pb}}\xspace}

\def\rfb   {\ensuremath{R_\mathrm{FB}}\xspace}
\def\rpa   {\ensuremath{R_{\pPb}}\xspace}

\def\Zmumu   {\ensuremath{\Z\to\mumu}\xspace}

\def\phistar {\ensuremath{\phi^{*}}\xspace}

\def\zrapstar {\ensuremath{y^{*}_{\Z}}\xspace}
\def\ZpT     {\ensuremath{p_{\mathrm{T}}^{\Z}}\xspace}

\def\powheg     {\mbox{\textsc{PowhegBox}}\xspace}
\usepackage{caption}
\usepackage{subcaption}
\begin{document}

\title{Probing the valence quark region of nucleons with \Z bosons at LHCb}

\author{\firstname{Hengne} \lastname{Li}\inst{1}\fnsep\thanks{\email{hengne.li@m.scnu.edu.cn}} \\ on behalf of the LHCb collaboration.}

\institute{Guangdong Provincial Key Laboratory of Nuclear Science, Guangdong-Hong Kong Joint Laboratory of Quantum Matter, Institute of Quantum Matter, South China Normal University, Guangzhou, China
}

\abstract{%
 In this high-$x$ region, both the flavour content and structure of the nucleon parton distribution functions remains relatively poorly known. 
New LHCb measurements of \Z and charm jet associated production could indicate a valence-like intrinsic-charm component in the proton wave function, and measurements of \Z production in \pPb collisions provide new constraints on the partonic structure of nucleons bound inside nuclei. 
Here we will discuss these new LHCb measurements and comparisons with state-of-the-art parton distribution function calculations.
}
\maketitle

LHCb is the only dedicated detector at LHC fully instrumented in the forward region with kinematic coverage $2<\eta<5$, allowing to study of the valence quark distributions of protons and nuclei in both small ($x < 10^{-3}$) and large ($10^{-1}<x<1$) Bjorken-$x$ regions.
The electroweak \Z-boson production and its leptonic decay products once produced do not participate in hadronic interactions. They are ideal for studying the non-perturbative initial-state effects.
Two recent results are presented in this proceedings, the study of the \Z boson and charm jets associated production~\cite{zcj} and the measurement of the \Z production in \pPb collisions~\cite{zmm}.

 Intrinsic charm (IC) refers to the long-timescale non-perturbative valence-like charm contents in protons, which is in contrast to the non-intrinsic charm contents arising from perturbative gluon radiation.
The \Z-boson production associated with charm jets probes intrinsic charm contents in protons.
Intrinsic charm contents in proton parton distribution functions (PDFs) can enhance charm jet production, especially at high Bjorken-$x$ when the production is associated with a \Z boson. The \Z production inherits a large momentum transfer $Q$ above the electroweak scale, implying at least one initial partons must have a sizeable Bjorken-$x$, corresponding to the high \Z-boson rapidity region.

\begin{figure}[t]
\centering
\includegraphics[width=0.45\columnwidth]{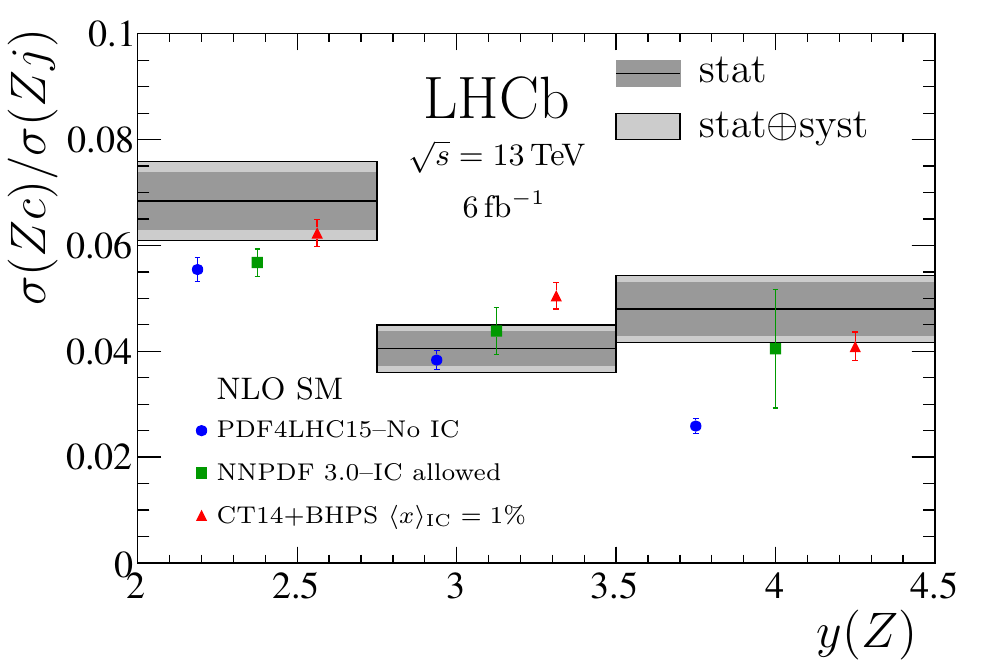}
\vspace*{-0.1cm}
\caption{Measured \Zcj distribution (grey bands) for three intervals of forward \Z rapidity,  compared to theoretical predictions.
}
\vspace*{-0.5cm}
\label{fig:zcj_results}
\end{figure}

The first measurement of the ratio of the production cross-sections 
$\Zcj \equiv \sigma(\Zc)/\sigma(\Zj)$ is reported by the LHCb experiment~\cite{zcj}, where \Zc refers to events containing a \Z boson and a charm jet, and \Zj refers to events containing a \Z boson and any type of jet.
The measurement takes the $pp$ collision samples collected by the LHCb detector from 2015 to 2018 corresponding to an integrated luminosity of about $6\invfb$.
The choice to measure the ratio \Zcj instead of $\sigma(\Zc)$ can largely cancel the experimental and theoretical uncertainties. The remaining leading systematic uncertainty is from the charm jet tagging.

The measured \Zcj in three \Z-boson rapidity intervals is shown in Fig.~\ref{fig:zcj_results}. Measurements are compared to next-leading-order (NLO) Standard Model (SM) predictions
without IC
, with the charm PDF shape allowed to vary (hence, permitting IC)
, and with IC as predicted by light front QCD calculations (referred to as the BHPS model) with a mean momentum fraction of 1\%
. 
A clear enhancement of the \Zcj can be observed in the highest \Z-boson rapidity interval, the inconsistency with respect to the no-IC expected value is greater than $3-\sigma$. 
More consistency with IC-allowed predictions can be seen, such as the BHPS model based on light front QCD.
The new high rapidity results should be able to strongly constrain the charm PDF in the large Bjorken-$x$ region.
Current results are statistically limited, Run3 dataset will allow for studies in finer \Z rapidity intervals.

\Z-boson production
can be used as clean probes of nuclear-matter effects on the initial state. 
The production of \Z bosons is sensitive to only the initial state, while hadronic probes are sensitive to both initial- and final-state nuclear matter effects. Therefore, together with hadronic probes, \Z-boson production can differentiate between the effects 
of the initial and final state.
Studies
also show that \Z-boson production in proton-lead (\pPb) collisions
at the \lhc is sensitive to heavier quark flavours. 
Improved information on the nuclear corrections is helpful to reduce proton PDF uncertainties and
essential for distinguishing the distributions for the different parton flavours.

A new measurement of the \Zmumu production in \pPb collisions at LHCb is presented~\cite{zmm}, where the differential cross-section,
the forward-backward ratio (\rfb) of the production cross-sections and the nuclear modification factors (\rpa) are measured for the first time as a function of the rapidity of the \Z boson in the centre-of-mass frame (\zrapstar),
the transverse momentum (\ZpT) and an angular variable \phistar.
The \pPb collision dataset used in this analysis was collected at $\sqsnn=8.16\tev$ in 2016 by the LHCb detector corresponding to an integrated luminosity of $12.2\pm0.3\invnb$ for forward collisions
and $18.6\pm0.5\invnb$ for backward collisions.

\begin{figure}[htbp]
\begin{center}
\begin{subfigure}[b]{0.32\textwidth}
\centering
\includegraphics[width=\textwidth]{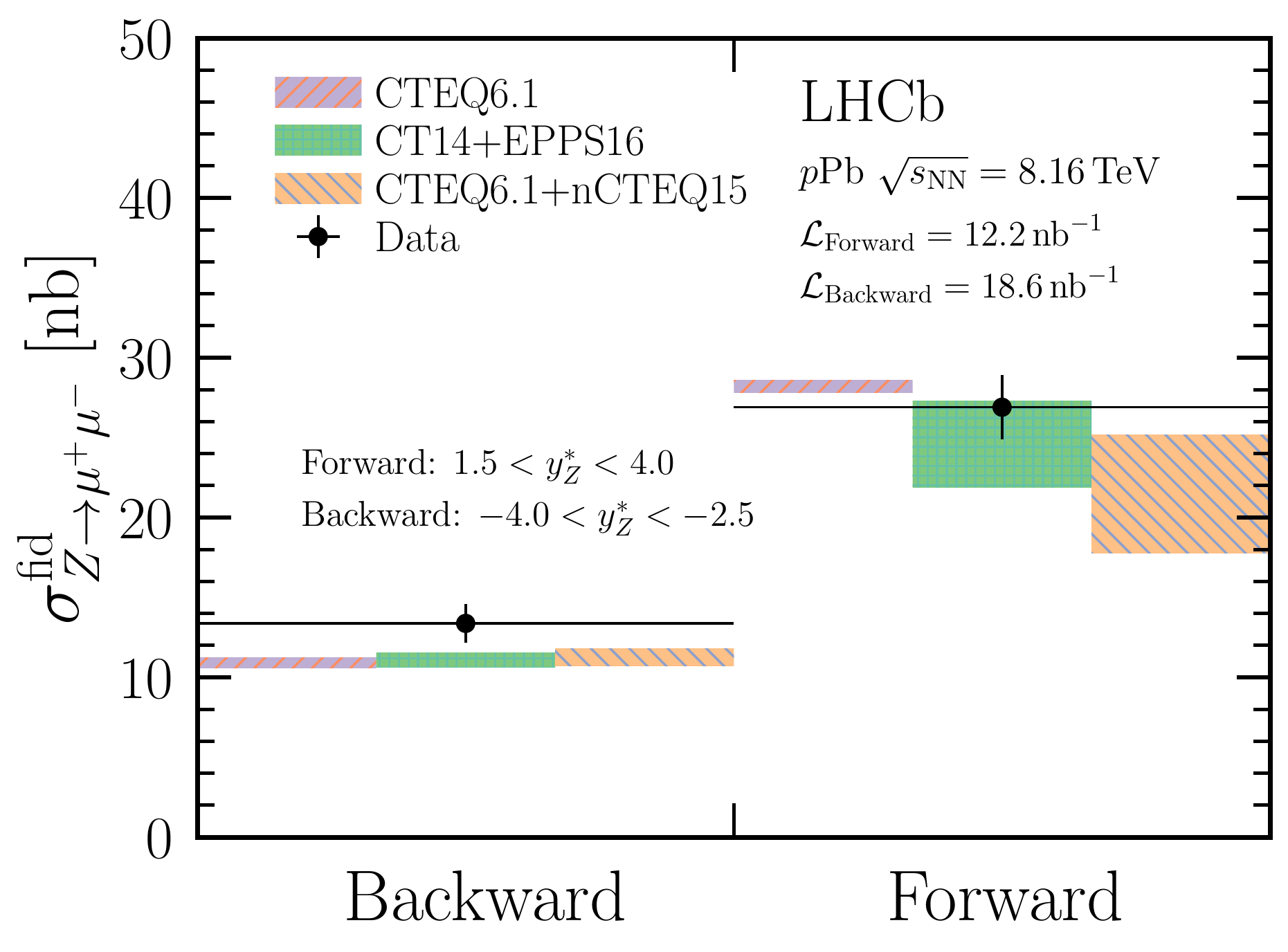}
\put(-30,20){(a)}
\end{subfigure}
\begin{subfigure}[b]{0.32\textwidth}
\centering
\includegraphics[width=\textwidth]{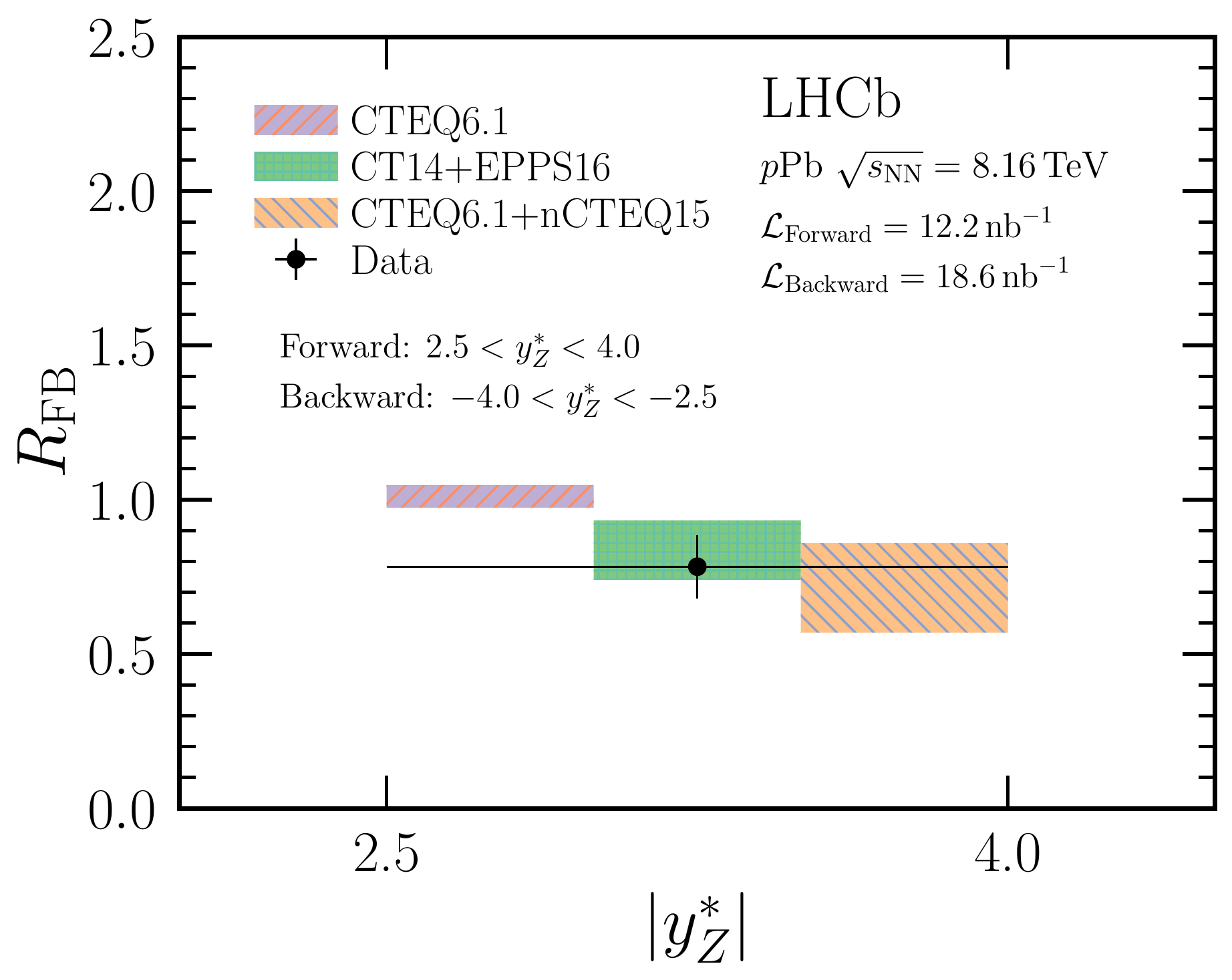}
\put(-30,25){(b)}
\vspace*{-.21cm}
\end{subfigure}
\begin{subfigure}[b]{0.32\textwidth}
\centering
\includegraphics[width=\textwidth]{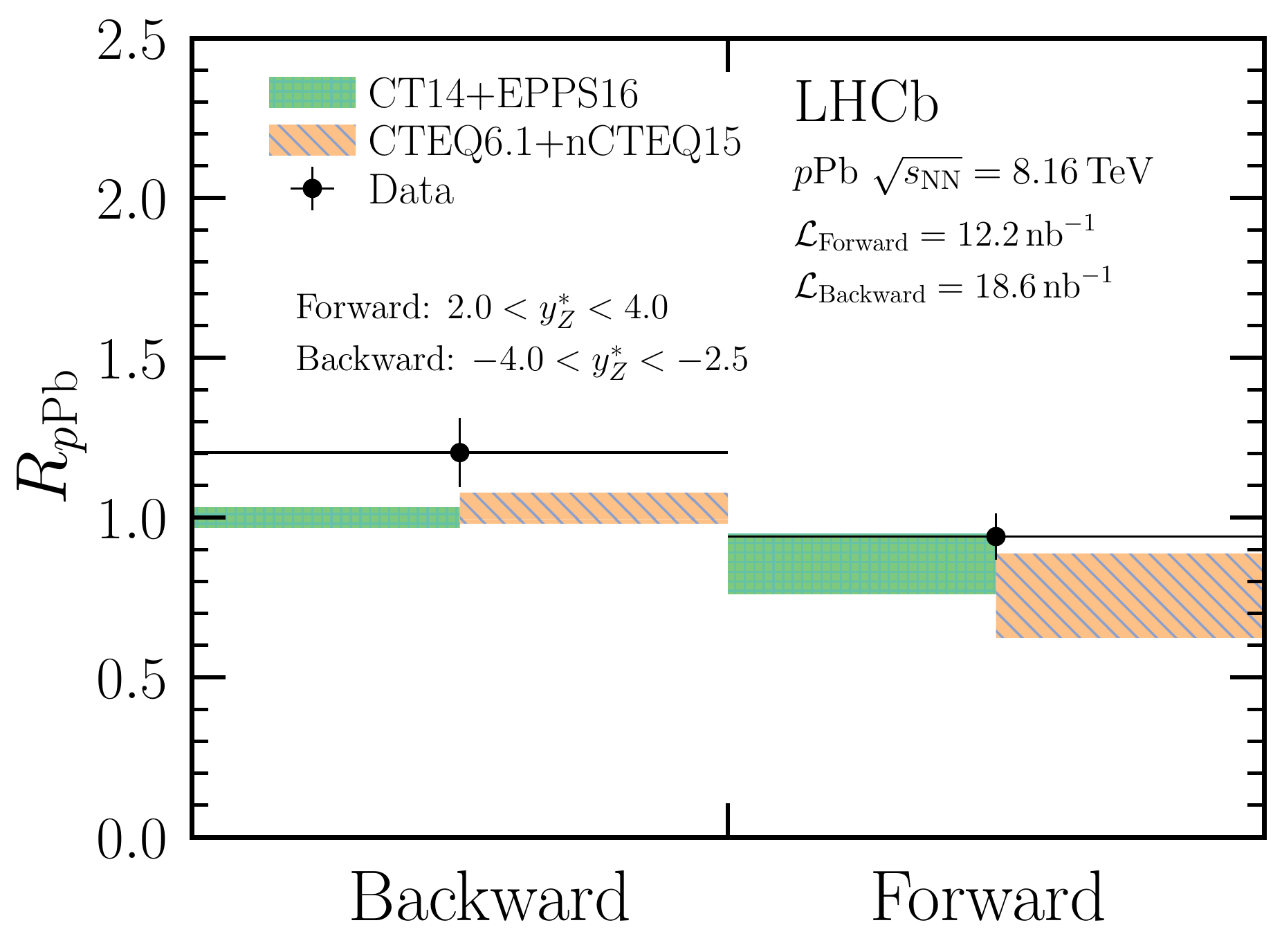}
\put(-30,20){(c)}
\end{subfigure}
\end{center}
\vspace*{-0.5cm}
\caption{
The measured inclusive (a) \Zmumu production fiducial cross-section, (b) forward-backward ratio (\rfb), and (c) nuclear modification factors (\rpa),  
compared to theoretical predictions.
}
\vspace*{-0.5cm}
\label{fig:zmmtotal}
\end{figure}

The measured inclusive fiducial cross-section, forward-backward ratio (\rfb), and nuclear modification factors (\rpa) are shown in Fig.~\ref{fig:zmmtotal}, together with
the comparisons to the \powheg prediction using
CTEQ6.1, EPPS16 and nCTEQ15 (n)PDF sets,
for forward and backward collisions, respectively.
For forward collisions,
the measured cross-section shown in Fig.~\ref{fig:zmmtotal}(a) appears to have a good agreement
with the \powheg calculations, with a smaller uncertainty
for the two intervals of $2.0<\zrapstar<3.0$ compared
to the theoretical calculations,
which can be used to further constrain the
nPDFs.
For backward collisions,
the uncertainty of the measurement is larger than
that of the \powheg calculation, and the measured central value is higher than the prediction especially
for the $-3.5 <\zrapstar <-3.0$ interval by about 2$\sigma$.
However, the measurement and calculation are compatible.
The measured value of \rfb shown in Fig.~\ref{fig:zmmtotal}(b) is below unity,
which is a reflection of the suppression
due to, {\it e.g.}, nuclear shadowing at small
Bjorken-$x$, together with an average
enhancement at large Bjorken-$x$.
The data is in agreement
with the EPPS16 and nCTEQ15 predictions.
The uncertainty of the measurement is
smaller than the theoretical uncertainties
using EPPS16 and nCTEQ15 nPDFs,
showing a constraining power on the
nPDFs.
The measured overall \rpa values are shown in Fig.~\ref{fig:zmmtotal}(c) and are 
compared to the \powheg predictions using the
EPPS16 and nCTEQ15 nPDF sets.
The overall \rpa results show good compatibility
between measurements and theoretical predictions.
The backward rapidity result shows
larger uncertainty compared to that of the
nPDF sets.
The measured central value is consistent with the
prediction at a 2$\sigma$ level.
The forward rapidity result
gives a higher precision than the
EPPS16 and nCTEQ15 nPDF sets,
and the central value is larger than
the prediction,
which shows a constraining power
on the current nPDF sets.

\begin{figure}[htbp]
\begin{center}
\begin{subfigure}[b]{0.305\textwidth}
\centering
\includegraphics[width=\textwidth]{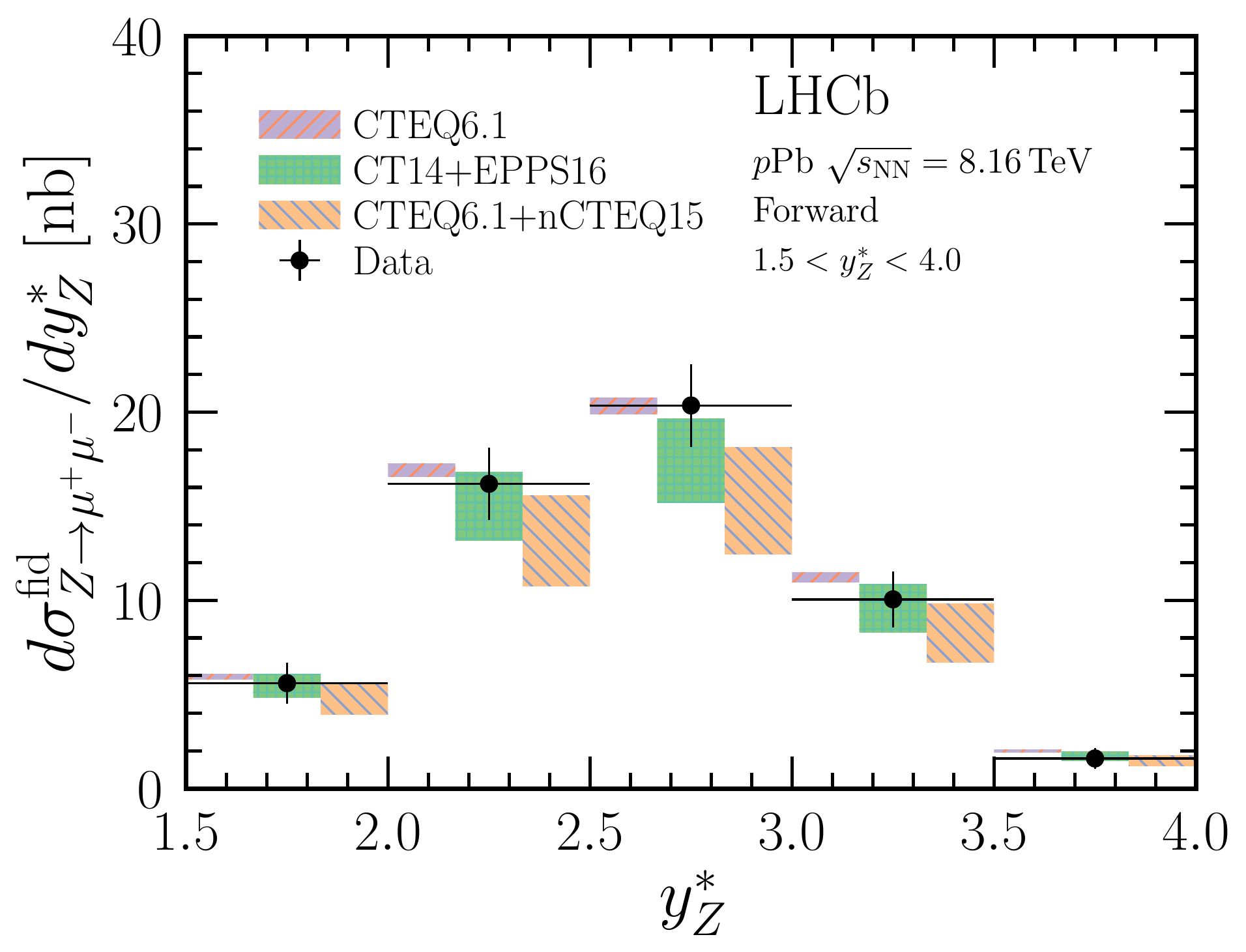}
\put(-25,50){(a)}
\end{subfigure}
\begin{subfigure}[b]{0.305\textwidth}
\centering
\includegraphics[width=\textwidth]{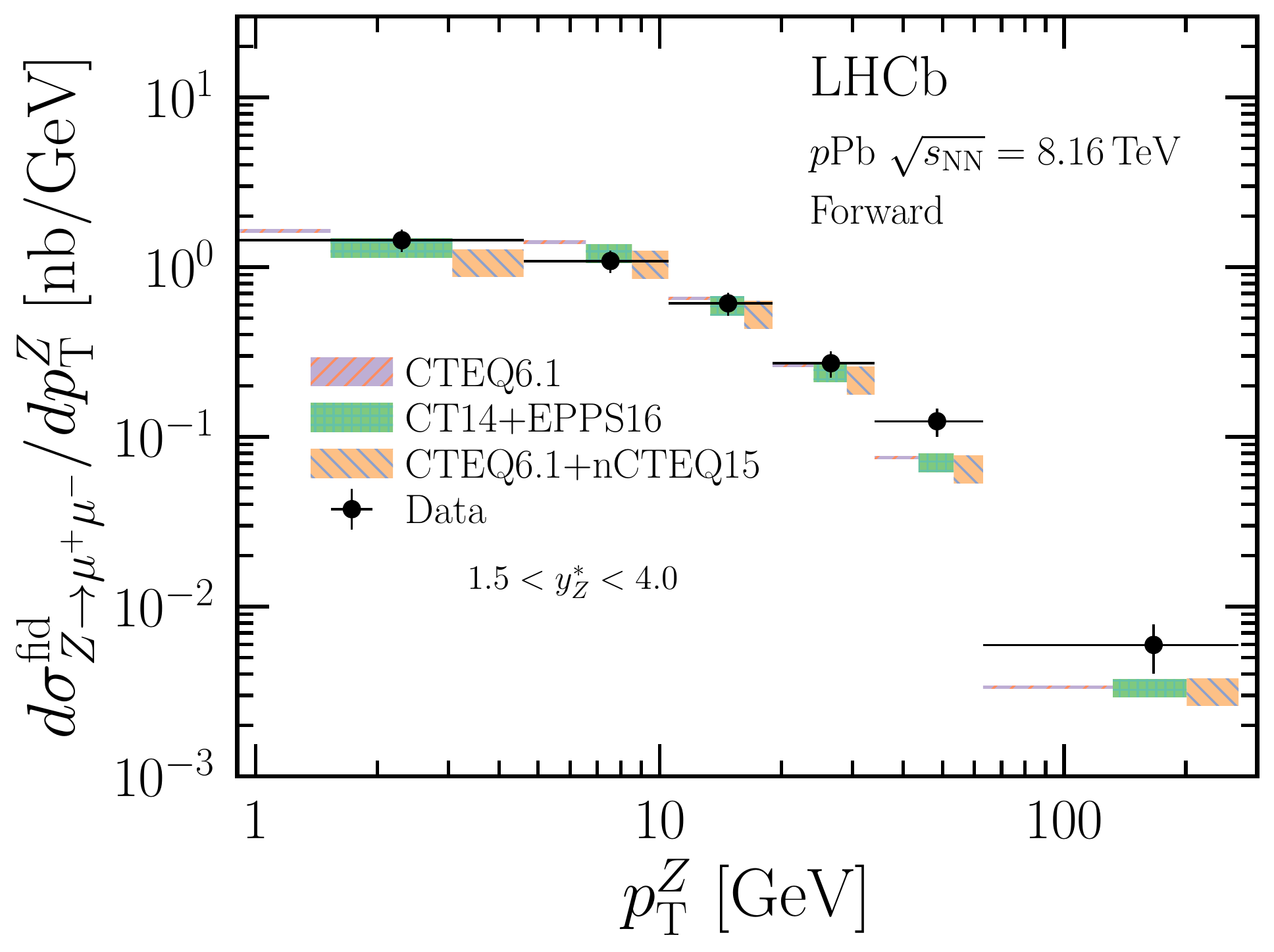}
\put(-25,50){(b)}
\vspace*{-.05cm}
\end{subfigure}
\begin{subfigure}[b]{0.32\textwidth}
\centering
\includegraphics[width=\textwidth]{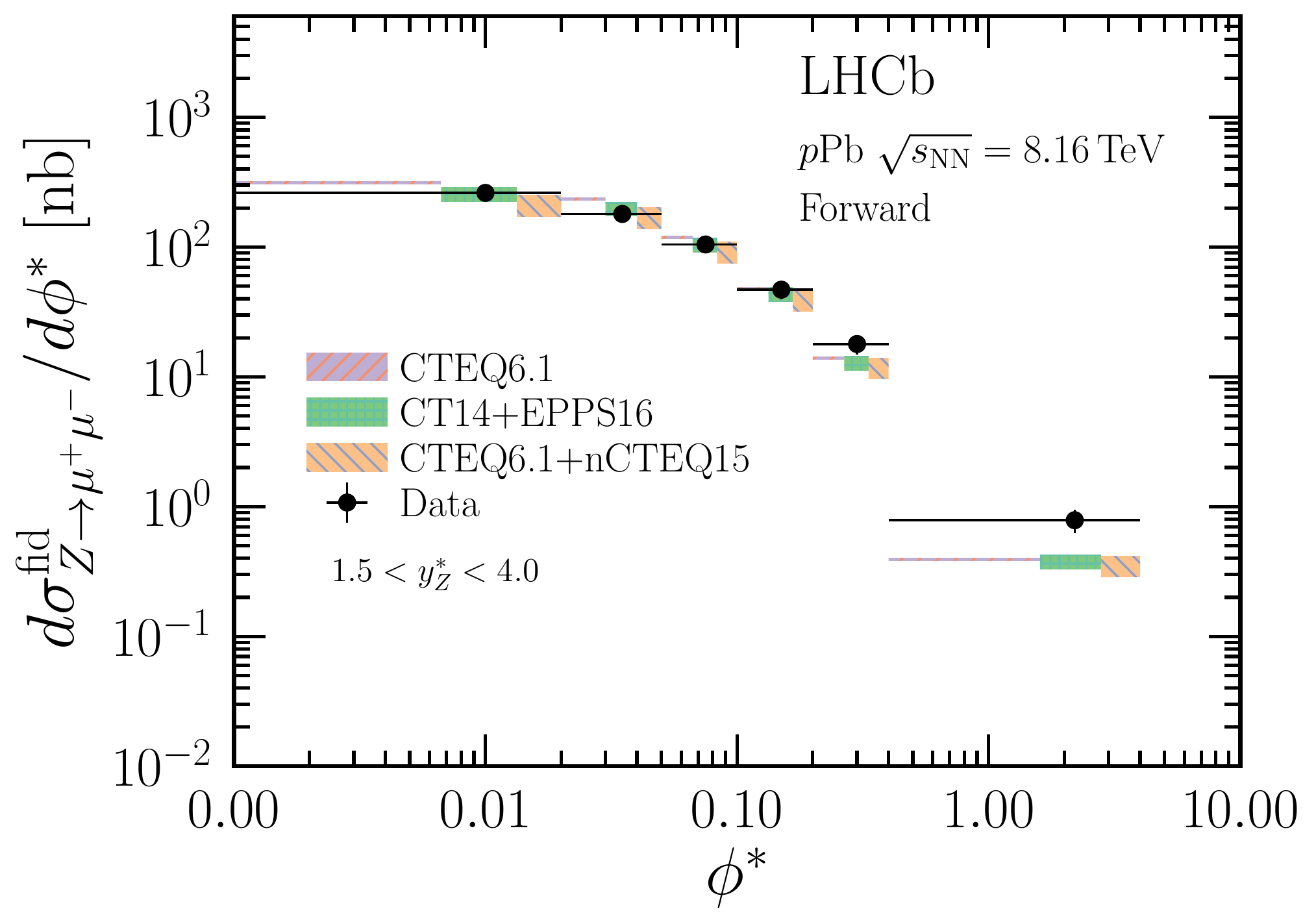}
\put(-25,50){(c)}
\end{subfigure}
\begin{subfigure}[b]{0.31\textwidth}
\centering
\includegraphics[width=\textwidth]{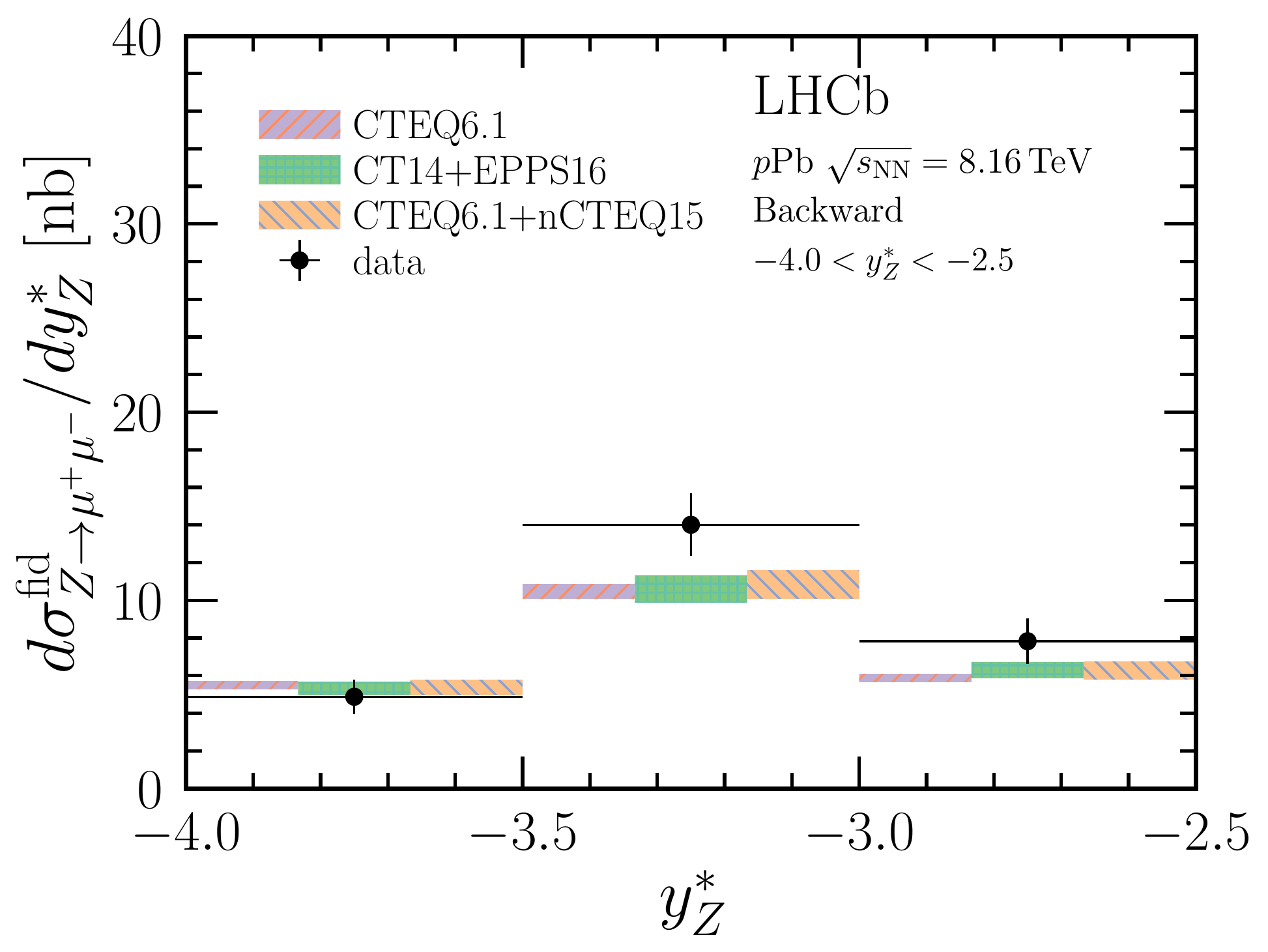}
\put(-25,50){(d)}
\end{subfigure}
\begin{subfigure}[b]{0.305\textwidth}
\centering
\includegraphics[width=\textwidth]{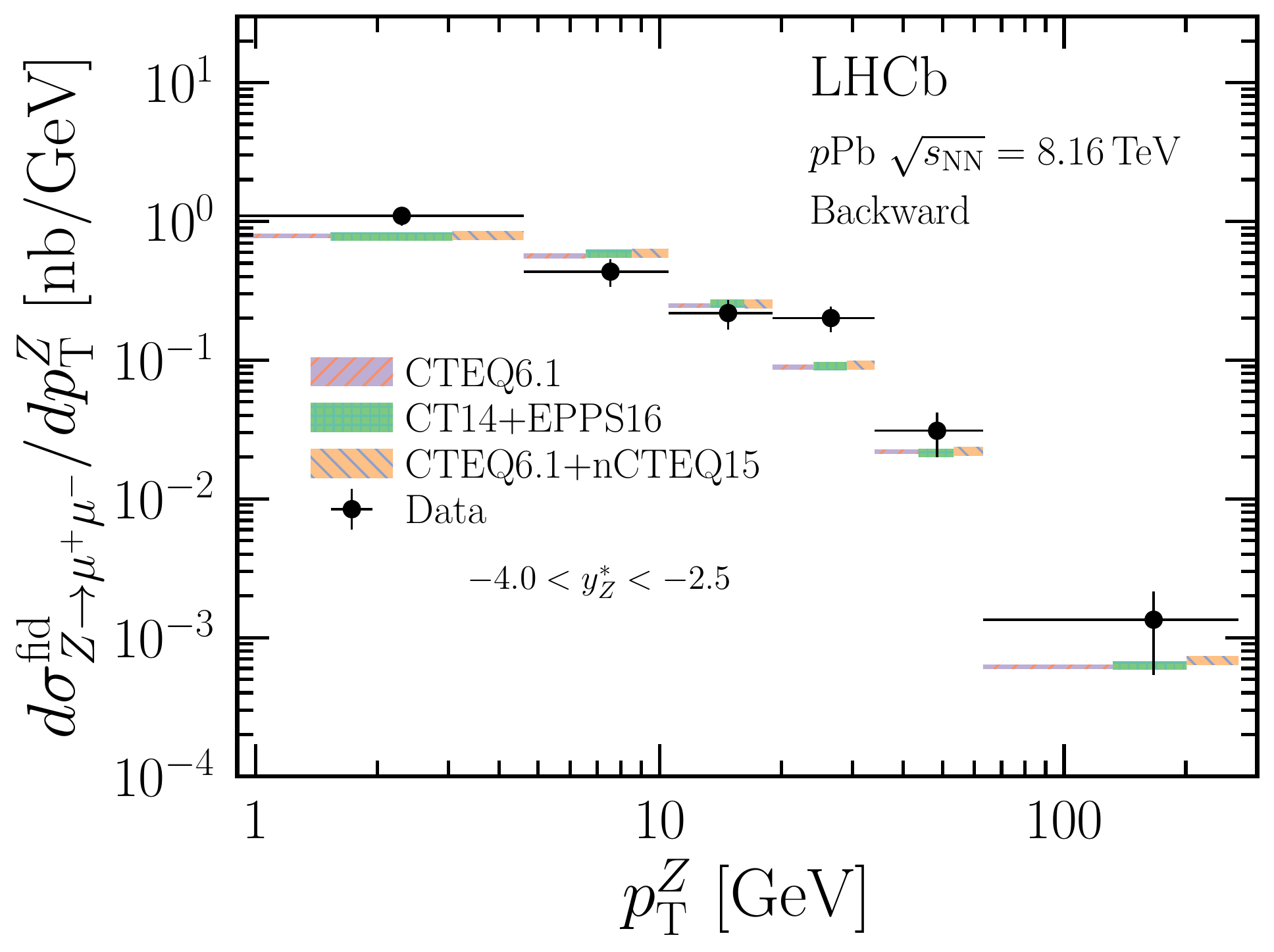}
\put(-25,50){(e)}
\vspace*{-.05cm}
\end{subfigure}
\begin{subfigure}[b]{0.32\textwidth}
\centering
\includegraphics[width=\textwidth]{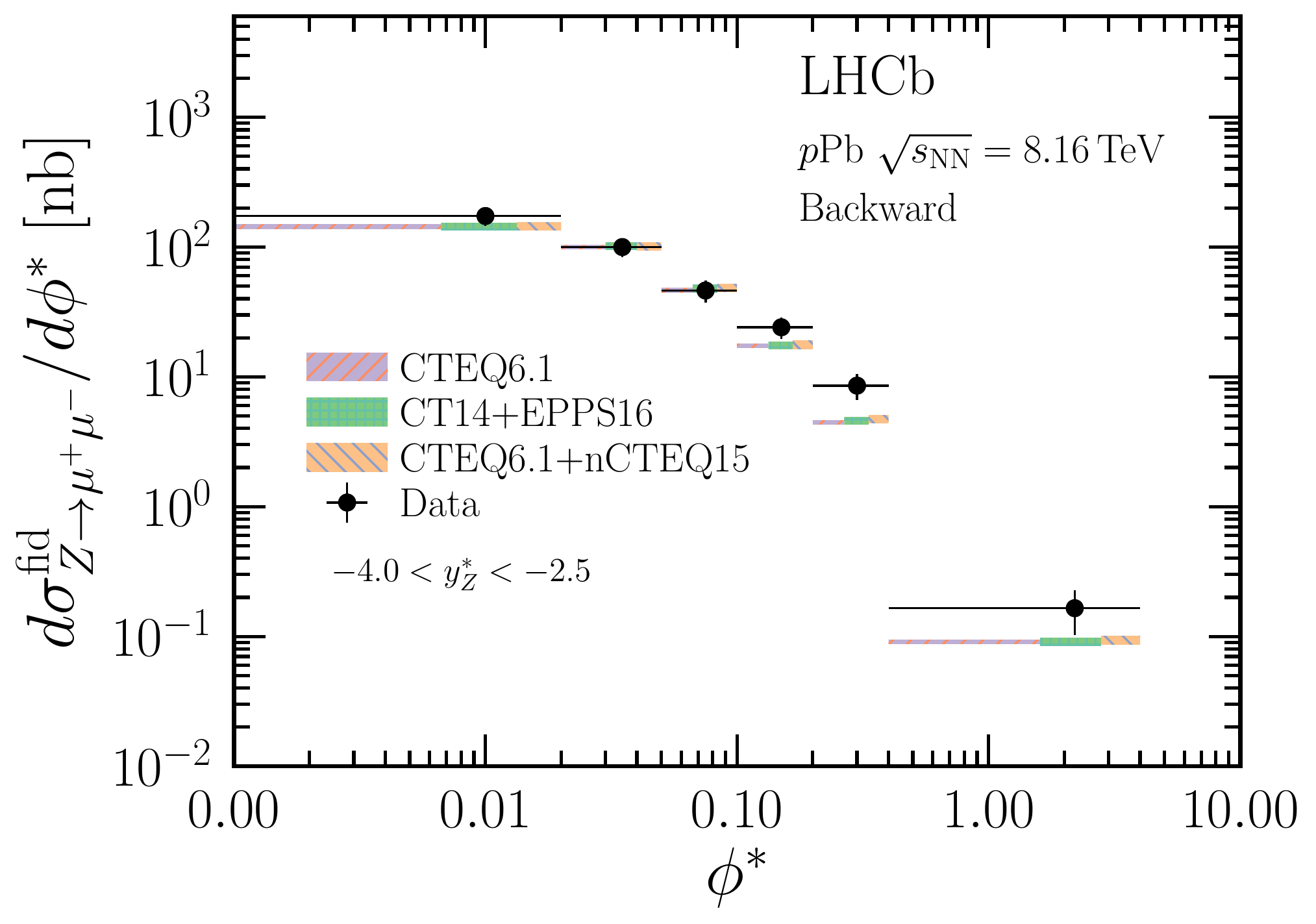}
\put(-25,50){(f)}
\end{subfigure}
\end{center}
\vspace*{-0.5cm}
\caption{
Differential cross-section as a function of (left column) \zrapstar, (middle column) \ZpT and (right column) \phistar,
together with the theoretical predictions,
where the top row is for forward collisions and the bottom row is for backward collisions.
}
\vspace*{-.5cm}
\label{fig:zmmxsec}
\end{figure}

\begin{figure}[htbp]
\begin{center}
\begin{subfigure}[b]{0.32\textwidth}
\centering
\includegraphics[width=\textwidth]{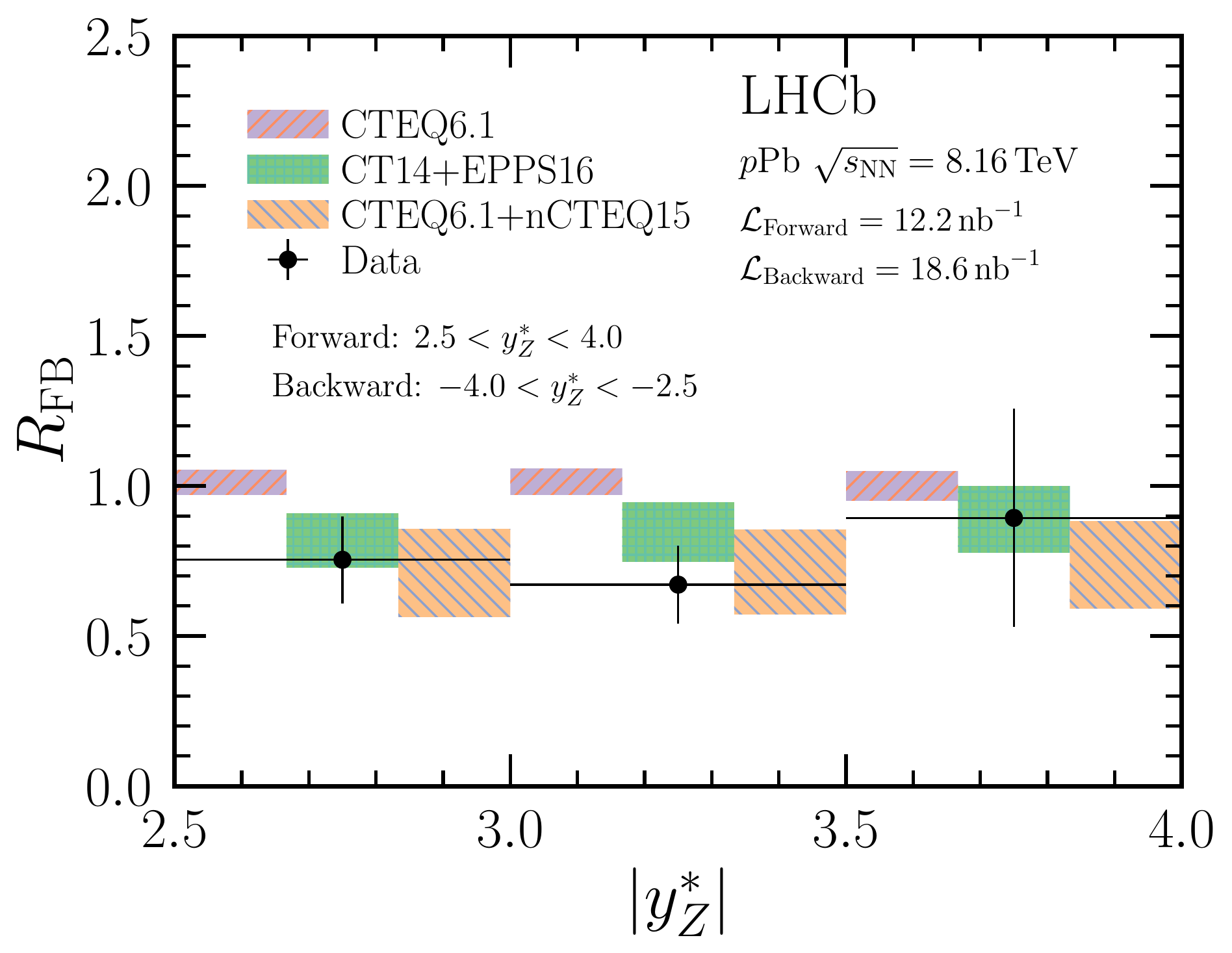}
\put(-40,50){(a)}
\end{subfigure}
\begin{subfigure}[b]{0.31\textwidth}
\centering
\includegraphics[width=\textwidth]{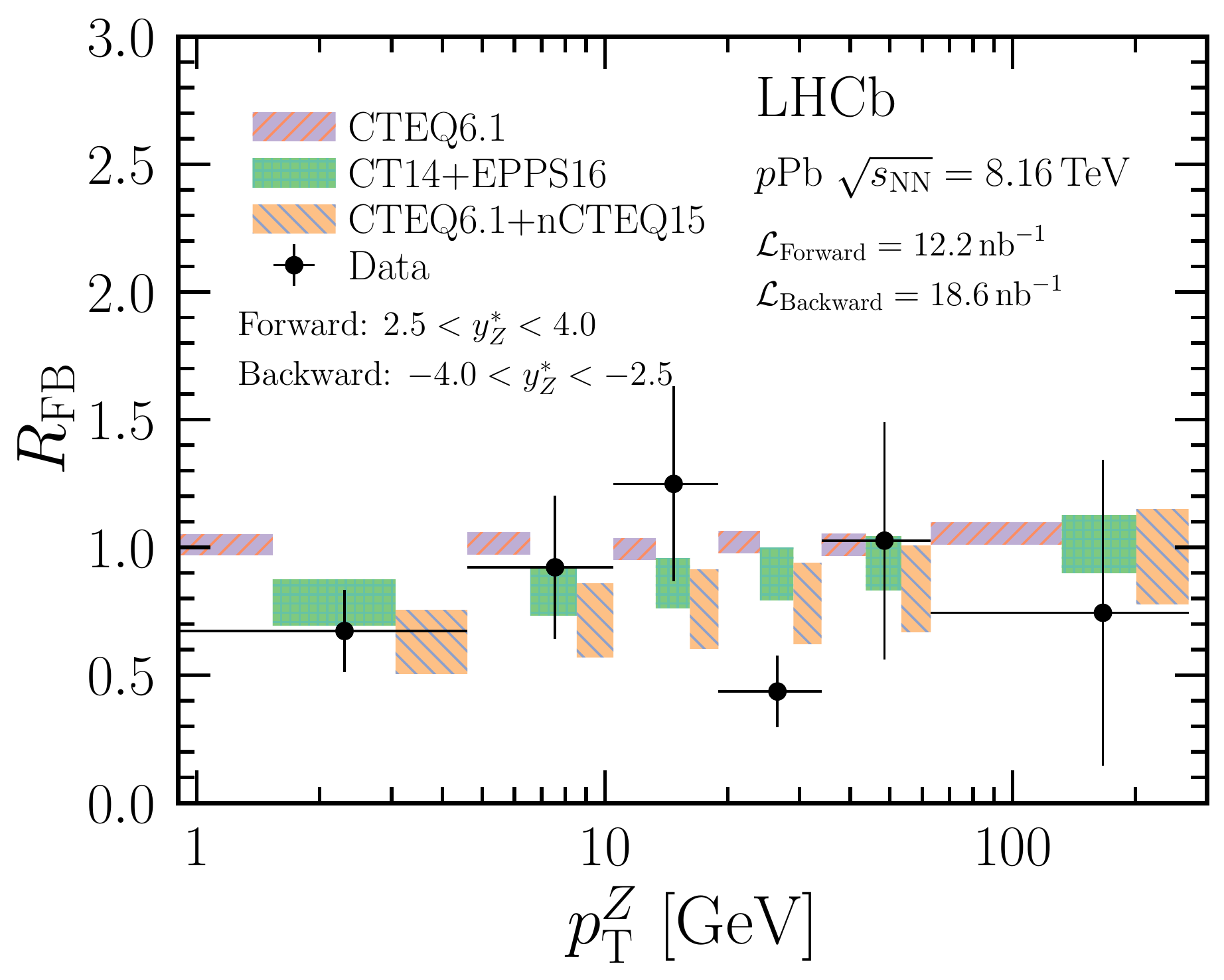}
\put(-40,52){(b)}
\vspace*{-.05cm}
\end{subfigure}
\begin{subfigure}[b]{0.32\textwidth}
\centering
\includegraphics[width=\textwidth]{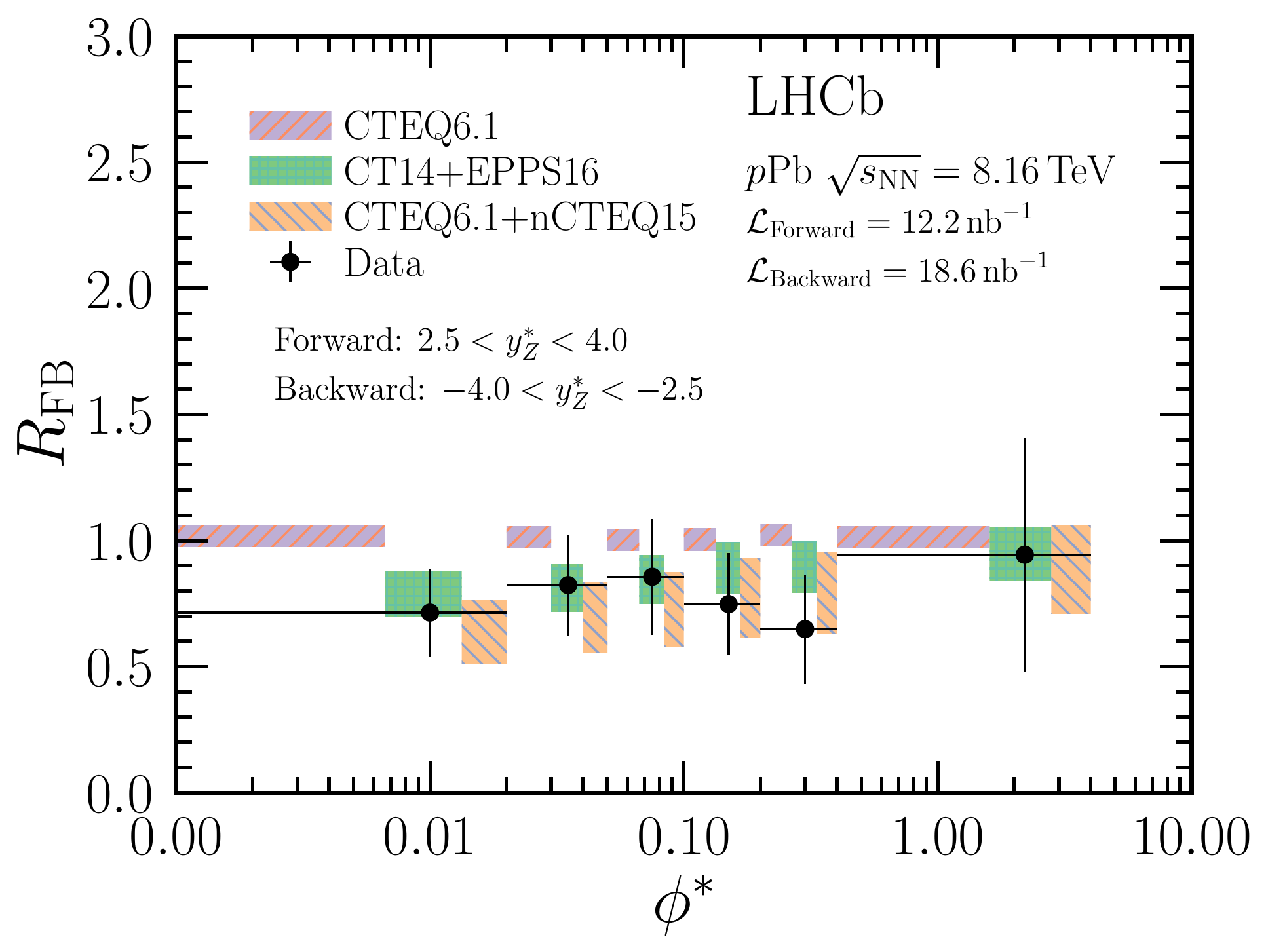}
\put(-40,50){(c)}
\end{subfigure}
\end{center}
\vspace*{-.5cm}
\caption{
Forward-backward ratio (\rfb) as a function of (left column) \zrapstar, (middle column) \ZpT and (right column) \phistar,
together with the theoretical predictions,
where the top row is for forward collisions and the bottom row is for backward collisions.
}
\vspace*{-.5cm}
\label{fig:zmmrfb}
\end{figure}

\begin{figure}[htbp]
\begin{center}
\begin{subfigure}[b]{0.32\textwidth}
\centering
\includegraphics[width=\textwidth]{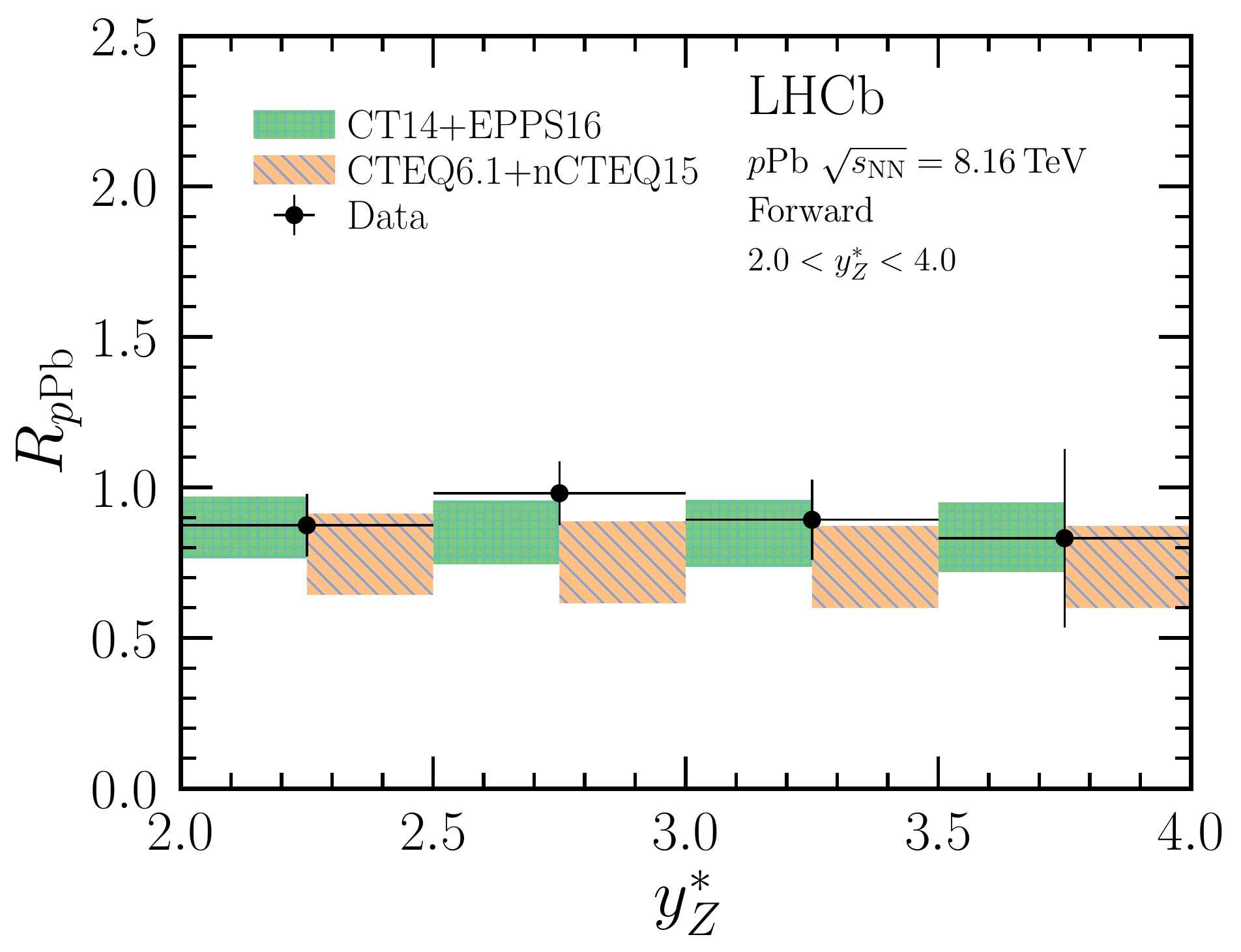}
\put(-30,20){(a)}
\end{subfigure}
\begin{subfigure}[b]{0.30\textwidth}
\centering
\includegraphics[width=\textwidth]{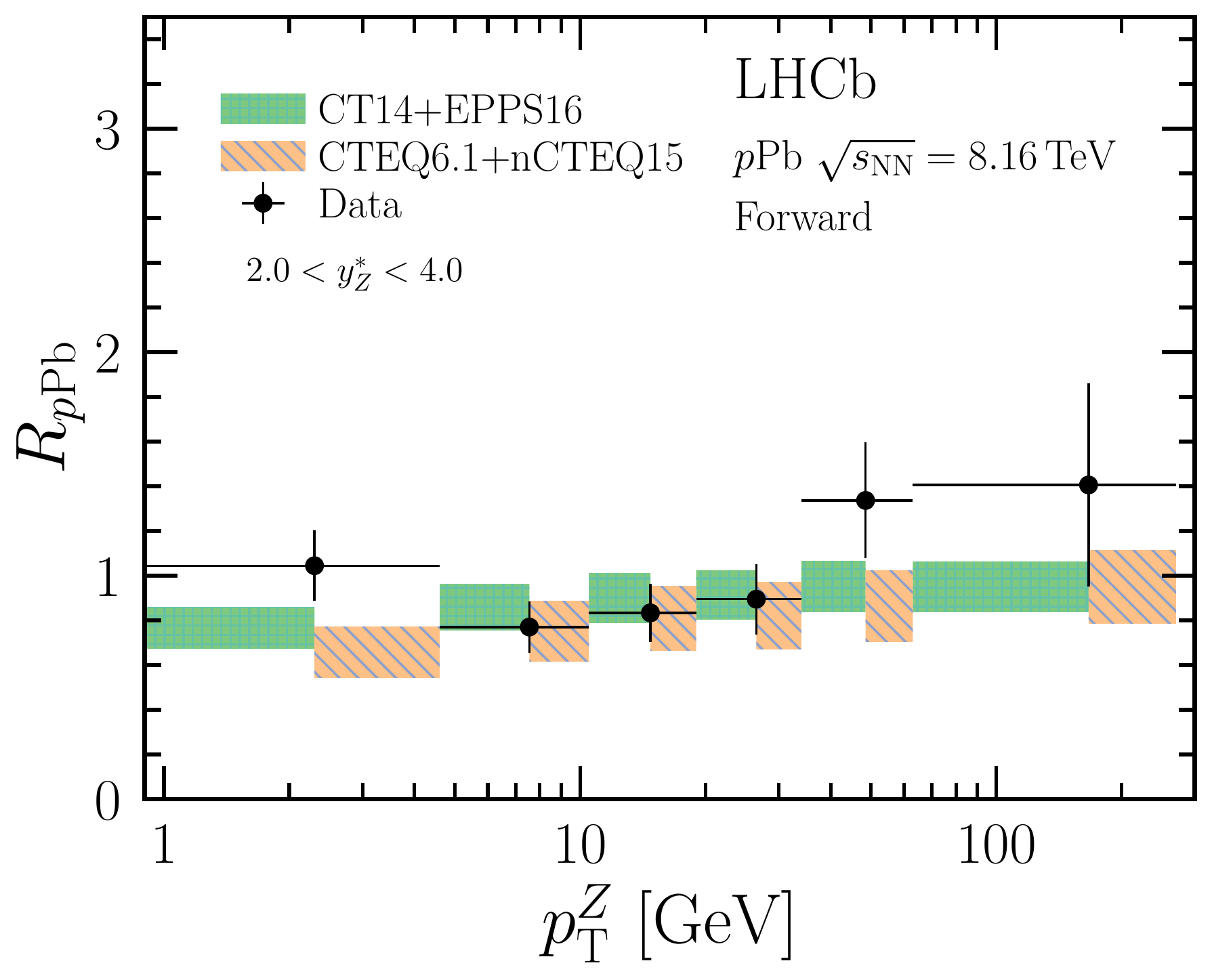}
\put(-30,25){(b)}
\vspace*{-.07cm}
\end{subfigure}
\begin{subfigure}[b]{0.32\textwidth}
\centering
\includegraphics[width=\textwidth]{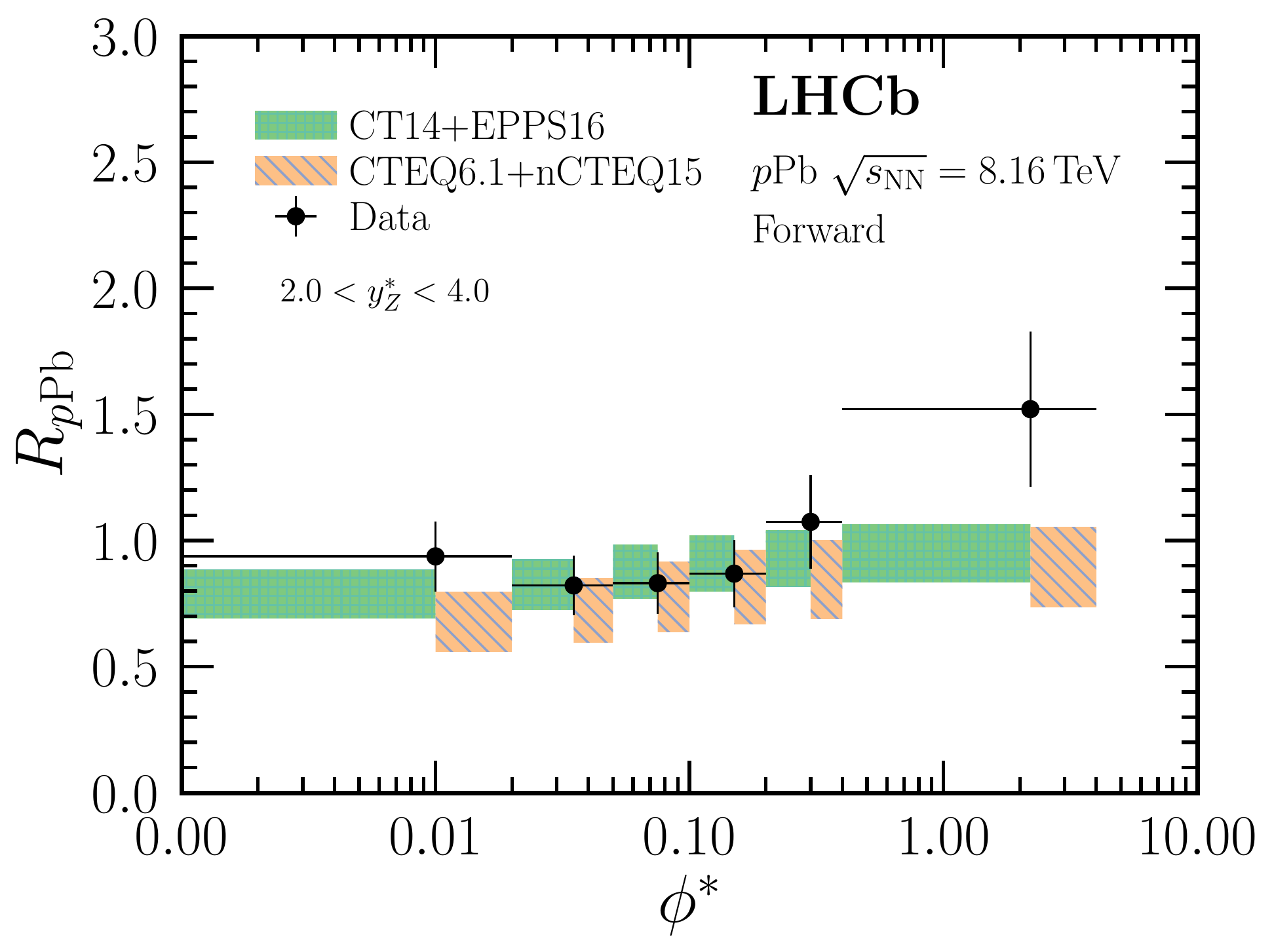}
\put(-30,20){(c)}
\end{subfigure}
\begin{subfigure}[b]{0.325\textwidth}
\centering
\includegraphics[width=\textwidth]{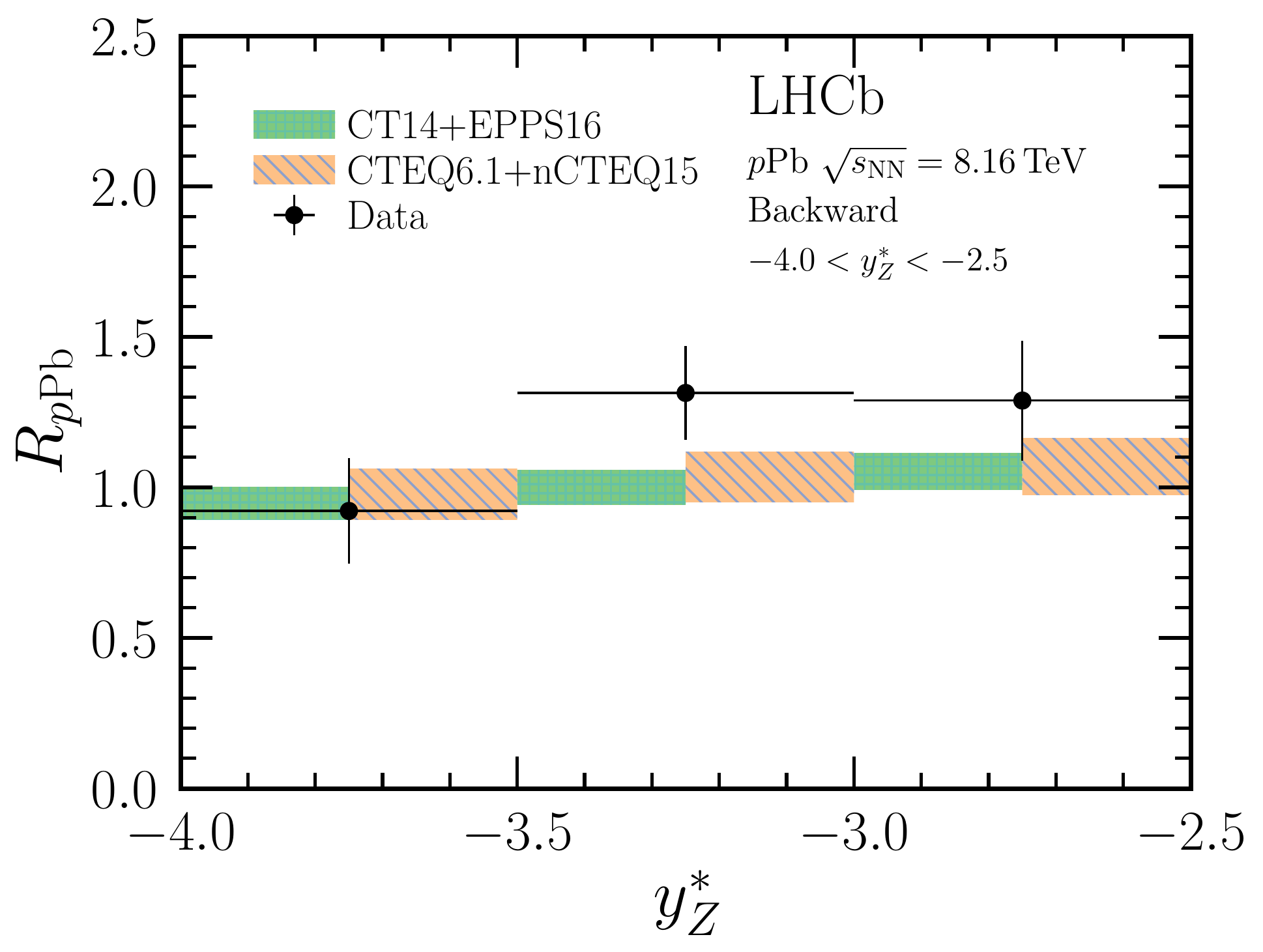}
\put(-30,20){(d)}
\end{subfigure}
\begin{subfigure}[b]{0.30\textwidth}
\centering
\includegraphics[width=\textwidth]{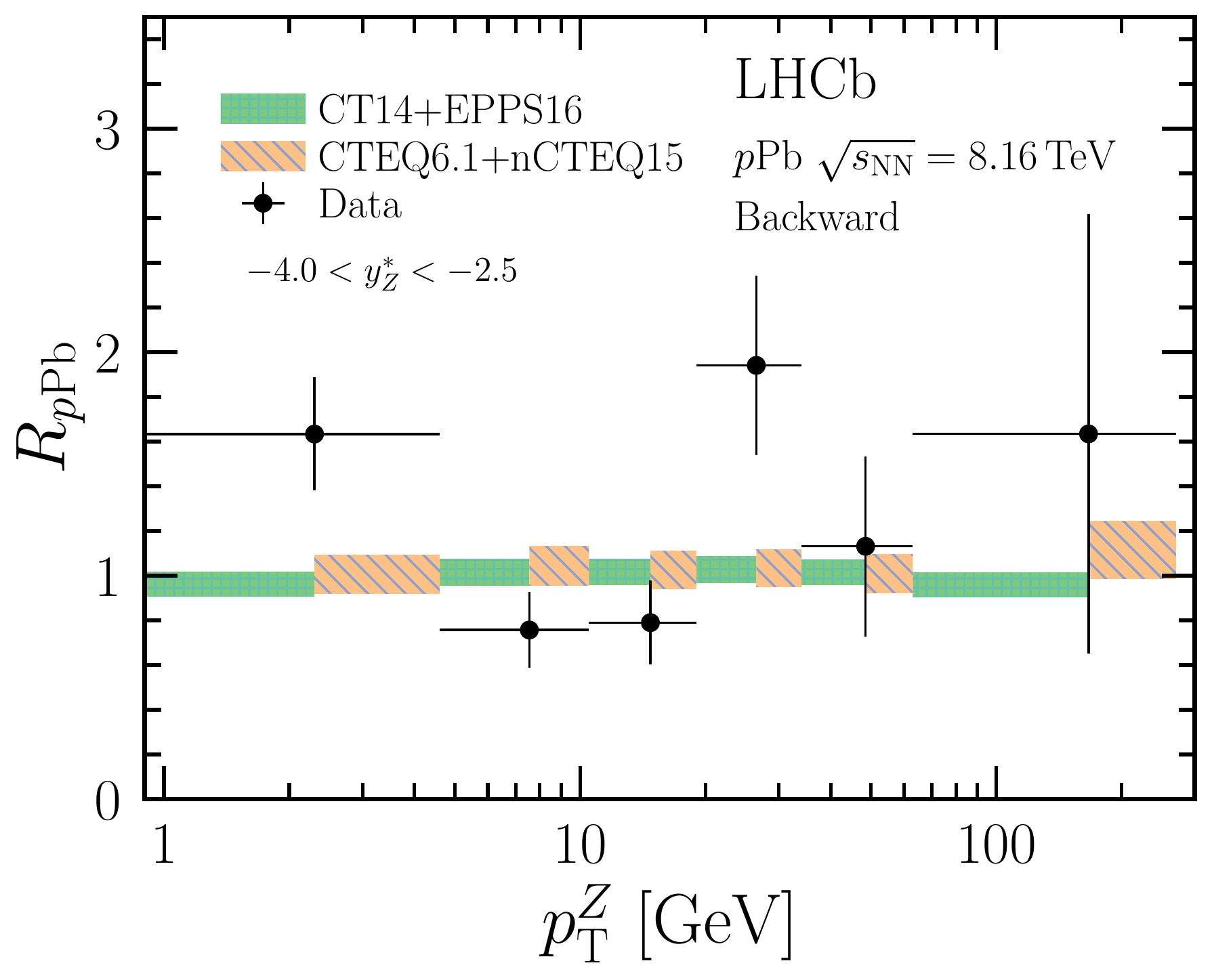}
\put(-30,25){(e)}
\vspace*{-.07cm}
\end{subfigure}
\begin{subfigure}[b]{0.32\textwidth}
\centering
\includegraphics[width=\textwidth]{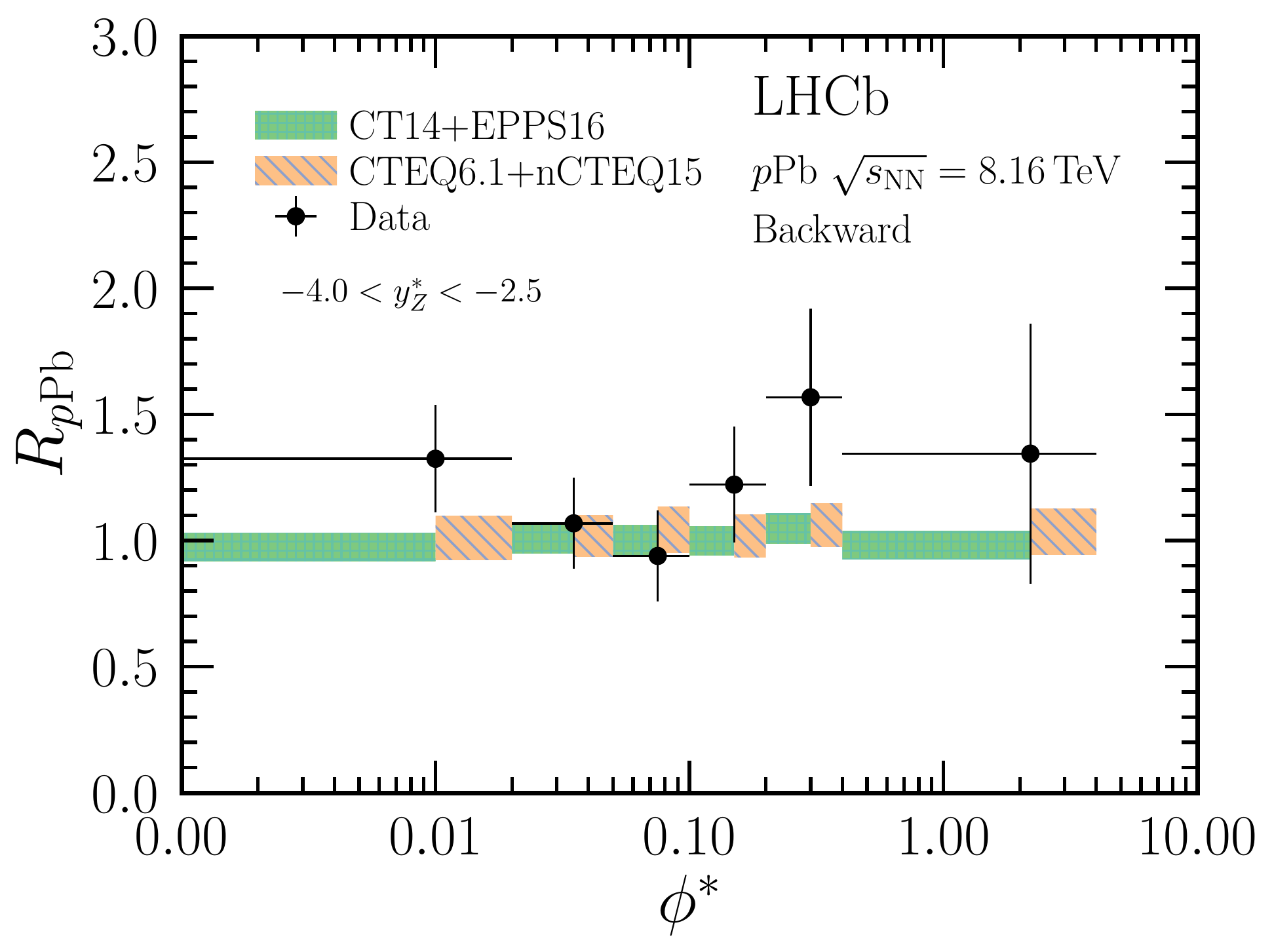}
\put(-30,20){(f)}
\end{subfigure}
\end{center}
\vspace*{-.5cm}
\caption{
Nuclear modification factors (\rpa) as a function of (left column) \zrapstar, (middle column) \ZpT and (right column) \phistar,
together with theoretical predictions,
where the top row is for forward collisions and the bottom row is for backward collisions.
}
\vspace*{-.5cm}
\label{fig:zmmrpa}
\end{figure}

The differential measurements of the production cross-sections, \rfb and \rpa are shown in Figs.~\ref{fig:zmmxsec}, \ref{fig:zmmrfb} and \ref{fig:zmmrpa}, respectively. 
In general, these results are compatible 
with nPDF predictions.
For the results of the cross-section and \rpa, larger uncertainties compared to the current nPDF predictions appear in the backward collisions for 
all three observables.
However, for forward rapidity
the large dataset gives a higher precision for certain intervals
compared to the nPDF predictions.
For the measured \rpa results, an overall suppression below unity as expected can be observed. However, given the current large uncertainties of the measurements, no conclusive statement can be made.

In summary, the electroweak \Z-boson production is an ideal probe of the initial conditions at hadron colliders. The fraction of \Z-boson production associated with charm jets is measured for the first time in the forward region at LHCb using $pp$ collisions at 13 TeV. 
Considerable enhancement is observed at the large \Z-boson rapidity, consistent with predictions assuming the existence of intrinsic (valence-like) charm contents.
A new \Z-boson production measurement in \pPb collisions at 8.16 TeV is reported.  
The differential cross-section, \rfb and \rpa as a function of \zrapstar, \ZpT, and \phistar are measured for the first time in the forward region at LHCb.
The new results are compatible with nCTEQ15 or EPPS16 nPDFs calculations.
Forward (small Bjorken-$x$) results show strong constraining power on the nPDFs.


\begin{thebibliography}{}
\bibitem{zcj}
LHCb collaboration, Phys. Rev. Lett. \textbf{128} (2022) 082001.
\bibitem{zmm}
LHCb collaboration, LHCb-PAPER-2022-009, arXiv:2205.10213 (2022), accepted by JHEP.
\end{thebibliography}
\end{document}